\documentclass[a4paper]{article}

\usepackage{geometry}
\usepackage{nameref}

\usepackage[figuresright]{rotating}
\usepackage{graphicx}

\usepackage{longtable}
\usepackage{array}
\usepackage{booktabs}

\usepackage{multirow}
\usepackage{makecell}

\usepackage{graphicx}
\usepackage{tensor}
\usepackage{mathtools}
\usepackage{amsmath}
\usepackage{amsxtra}

\usepackage{upgreek}

\usepackage{appendix}

\usepackage{bm}

\usepackage[noblocks]{authblk}

\usepackage[numbers,sort&compress]{natbib}

\usepackage{CJK}
\usepackage{txfonts}
\usepackage{times}
\usepackage{ragged2e}
\usepackage{indentfirst}

\usepackage{verbatim}

\usepackage{ntheorem}

\usepackage[colorlinks=true,linkcolor=blue,citecolor=blue,urlcolor=green]{hyperref}

\renewcommand{\raggedright}{\leftskip=0pt \rightskip=0pt plus 0cm}
\usepackage{setspace}
\allowdisplaybreaks[4]

\begin{document}

\title{Security proof for quantum cryptography against entanglement-measurement attack}

\author{Zhaoxu Ji$^\dag$, Huanguo Zhang
\\
{\small {Key Laboratory of Aerospace Information Security and Trusted Computing, 
Ministry of Education, School of Cyber Science and Engineering, Wuhan University, Wuhan 430072 China\\
$^\dag$jizhaoxu@whu.edu.cn}}}
\date{}


\maketitle


\begin{abstract}

Entanglement-measurement attack is one of the most famous attacks against quantum cryptography.
In quantum cryptography protocols, eavesdropping checking is an effective means to resist this attack.
There are currently two commonly used eavesdropping checking methods: one is to prepare two sets of non-orthogonal 
single-particle states as decoy states, and determine whether there are eavesdroppers in the quantum channel 
by comparing the states obtained by measurements with the original states;
The other is to prepare two sets of non-orthogonal entangled states and use their entanglement correlations to 
judge whether there are eavesdroppers in the quantum channel.
In this paper, we theoretically demonstrate how quantum cryptography can
utilize these two eavesdropping checking methods to resist entanglement-measurement attacks.
We take the quantum cryptography protocols based on maximally entangled states as examples to 
demonstrate the proof process, transitioning from qubit-based protocols to qudit-based ones.

\end{abstract}


\noindent
\textbf{Keywords}: quantum cryptography, quantum secure multi-party computation, security proof, 
entanglement-measurement attack


\section{Introduction}

\noindent
Quantum mechanical phenomena, such as quantum superposition and entanglement, are not difficult to talk about, 
and their basic formulations are not very complicated \cite{JiZX202002555}. 
However, the fundamental reasons for the existence of these phenomena have yet to be explained 
by the scientific community in a universally convincing manner,
indicating that it is difficult to obtain ideal answers solely through scientific exploration \cite{JiZX12442025}.
In fact, ancient Chinese philosophers (e.g. Confucius and Lao Zi) had already gained insight into some of the mysteries of 
the universe and provided wise perspectives \cite{JiZX12442025}. Unfortunately, in this era,
people who are interested in ancient Chinese philosophy, especially those who truly understand it, are extremely rare.
In fact, scientists have not been able to prove that the core teachings of mainstream religions are wrong, 
and many inexplicable bizarre phenomena, such as quantum mechanical phenomena, 
to some extent, support the correctness of certain core religious views.
If a person absolutely believes in science while completely denying religious teachings, the harm is self-evident.

Quantum mechanics has not only brought about innovation in people's ideological concepts,
but also brought new opportunities and challenges to multiple fields such as chemistry, atomic physics, 
information security, and computational science. Quantum information science,
formed by the combination of quantum mechanics and information science,
has become a key force leading a new round of technological revolution and 
industrial transformation \cite{ZhangHG16102019,GalindoA7422002}.
Major countries in the world have regarded it as a priority development direction for strategic technological strength.

As a major component of quantum information science, quantum cryptography is renowned worldwide 
for its unconditional security and eavesdropping checking capabilities \cite{ZhangHG16102019}. 
The earliest idea of quantum cryptography was proposed by Wiesner in the late 1960s and early 1970s, 
and later developed by Bennett and Brassard, who designed the first quantum key distribution protocol in 1984, 
(i.e., the famous BB84 protocol). Quantum cryptography has attracted widespread attention 
from researchers in fields such as physics, mathematics, and computer science, 
and has been applied in many countries to the sectors with high demand for information security, such as military and banking.
At present, the research scope of quantum cryptography is considerably extensive,
including many research directions such as quantum key distribution, quantum key agreement, 
quantum secure direct communication, quantum signature,
quantum authentication, and quantum secure multi-party computation \cite{ZhangHG16102019}.

Although significant progress has been made in both theoretical research and practical applications of quantum cryptography, 
the development of quantum cryptography still faces many serious challenges \cite{NielsenMA2000,JiZX5852022}. 
On the one hand, as one of the core resources of quantum cryptography, 
quantum entangled states are very difficult to prepare with existing quantum technology, especially multi-particle entangled states.
On the other hand, quantum systems are highly susceptible to the interference from the external environment, 
resulting in changes in the state, and thus the preservation and storage of quantum states is very challenging.
Furthermore, there are various attack methods against quantum cryptography protocols, 
and it is almost impossible to theoretically prove that a quantum cryptography protocol can resist all types of attacks. 
As a result, a large number of quantum cryptography protocols have already been shown to 
have information leakage issues \cite{ZhangHG16102019}.

Among the attack methods targeting quantum cryptography protocols, 
entanglement-measurement attack is one of the most well-known \cite{JiZ962021,JiZX72019,ZXJi1862019}.
To steal users' information, an eavesdropper prepares some auxiliary particles in advance. 
After intercepting particles in the quantum channel, the eavesdropper performs certain unitary operations to 
entangle the auxiliary particles with the intercepted particles. 
Finally, the eavesdropper performs quantum measurements on the auxiliary particles to 
steal the information carried by the intercepted particles.
It can be seen that this attack does not directly measure the intercepted particles, 
hence it theoretically avoids changing the states of these particles and has a high degree of confidentiality,
which poses a great threat to the security of quantum cryptography protocols.
Fortunately, quantum cryptography can effectively resist this attack through eavesdropping checking.
Eavesdropping checking is one of the most remarkable advantages of quantum cryptography
over classical cryptography, and it is also a unique feature of quantum cryptography.

There are currently two commonly used methods for eavesdropping checking, 
both of which have a major common feature: they require the preparation of two sets of non-orthogonal quantum states. 
The first method originated from the BB84 protocol, in which protocol participants 
prepare two sets of non-orthogonal single-particle basis, then measure them after completing transmission, 
and finally compare the quantum states obtained by measurements with the initial states to 
determine whether there is an eavesdropper in the quantum channel \cite{JiZ962021,JiZX72019,ZXJi1862019}. 
Another method is for protocol participants to first prepare two sets of non-orthogonal entangled states, 
then measure them after transmission is completed, and finally determine whether there are eavesdroppers 
in the quantum channel by checking the entanglement correlation of the quantum states obtained by 
measurements \cite{HuangW8932014,WangQ9422016}.
As is well known, preparing single-particle states with current technology is more feasible than preparing entangled states, 
hence most quantum cryptography protocols adopt the first method. Of course, 
with the continuous advancement of quantum technology, the feasibility of the second method will become increasingly high.

In fact, the security proof against the entanglement-measurement attack varies in form with different quantum states. 
However, the existing quantum states are diverse, hence it is very difficult, even almost impossible, 
to systematically provide a security proof for quantum cryptography to resist this attack,
which is probably the reason why there is no literature devoted to the study of this problem so far.
In this context, we focus on this problem in this paper. Although we cannot provide a systematic security proof, 
we can attempt to provide a basic framework for the proof, which has guiding significance for the security proof 
of quantum cryptography protocols based on different types of quantum states.
We assume that the quantum states used in the protocol are all pure states,
and transition from the protocols based on 2-level quantum systems (qubit-based protocols)
to the protocols based on d-level quantum systems (qudit-based protocols).
For the protocols that utilize the entanglement correlation of entangled states to resist entanglement-measurement attack, 
we consider maximally entangled states, which are the most commonly used and representative entangled 
states in quantum cryptography.

The rest of this paper is structured as follows. In Sec. 2, we first give the security proof of quantum cryptography protocols
using single-particle states to resist entanglement-measurement attack,
then we prove the security for the protocols using the entanglement correlations of entangled states to resist the attack. 
We conclude this paper in Sec. 3.


\section{Security proof against entanglement-measurement attack}

\noindent
The entanglement-measurement attack can be briefly described as follows \cite{JiZ962021,JiZX72019}:
When a participant (assumed to be Alice) in the quantum cryptography protocol sends the particles as information carriers 
to another participant (assumed to be Bob),
an eavesdropper (conventionally called Eve) intercepts part or all of the particles through quantum channels,
and entangles them with the ancillary particles that she prepares beforehand.
Then, she sends the particles with auxiliary particles to Bob.
Finally, Eve performs quantum measurements on the ancillary particles at the appropriate time,
such as after the protocol ends, to steal useful information.

\subsection{Security proof of using single-particle states for eavesdropping checking}

In quantum cryptography protocols, the two sets of non-orthogonal single-particle basis states 
(often called decoy particles or decoy photons in quantum cryptography) commonly used for eavesdropping checking
are $\{ \left\vert k \right\rangle \}_{k=0}^{d-1}$ and $\{ F \left\vert k \right\rangle \}_{k=0}^{d-1}$,
where $F$ represents the quantum Fourier transform,
\begin{equation}
\label{Fourier_transform}
	F \left\vert k \right\rangle = 
	\frac{1}{\sqrt d} \sum_{r=0}^{d-1} \zeta^{kr} \left\vert r \right\rangle, \quad \zeta = e^{2\pi i/d}.
\end{equation}
When $d = 2$, one can get the two state sets $\{\left|0\right\rangle,\left|1\right\rangle\}$
and $\{\left\vert + \right\rangle,\left\vert - \right\rangle\}$, which are widely used for eavesdropping checking 
in qubit-based quantum cryptography protocols.

In what follows, we would like to take the quantum cryptography protocol based on Bell states 
(also called EPR pairs \cite{BellJS131964})
as an example to demonstrate the security proof against the entanglement-measurement attack,
the basic form of which was presented in Ref. \cite{JiZX72019}.
By the way, the reason we chose the Bell-state-based quantum cryptography protocols is that
Bell states are one of the fundamental and most widely used entangled states as information carriers.

Without losing generality, it is assumed that Alice sends the particles in all or part of the Bell States to Bob 
(whether she sends all the Bell states or only some of them is determined by the protocol procedure),
and marks the sequence formed by these particles as $S$.
For eavesdropping checking, Alice prepares some decoy particles, each of which is randomly selected from the set
$\{\left|0\right\rangle,\left|1\right\rangle\}$ and $\{\left|+\right\rangle,\left|-\right\rangle\}$.
Then Alice inserts these decoy particles into the particle sequence $S$ at random positions and 
labels the newly generated sequence as $S^*$.
At this time, the eavesdropper Eve intercepts all or part of the particles in $S^*$ through the quantum channel,
and entangles the auxiliary particles on them, in which the entanglement action can be expressed as \cite{JiZ962021,JiZX72019}
\begin{gather}
U \left|0\right\rangle \left|\varepsilon\right\rangle =
\lambda_{00} \left|0\right\rangle \left| \epsilon_{00} \right\rangle 
+ \lambda_{01} \left|1\right\rangle \left| \epsilon_{01} \right\rangle,
\notag \\
U \left|1\right\rangle \left|\varepsilon\right\rangle =
\lambda_{10} \left|0\right\rangle \left| \epsilon_{10} \right\rangle 
+ \lambda_{11} \left|1\right\rangle \left| \epsilon_{11} \right\rangle,
\label{U-action-case-one}
\end{gather}
where $U$ is a unitary operation, $\left|\varepsilon\right\rangle$ is an ancillary particle,
$\left\vert \epsilon_{00} \right\rangle$, $\left| \epsilon_{01} \right\rangle$,
$\left\vert \epsilon_{10} \right\rangle$, $\left\vert \epsilon_{11} \right\rangle$
are the states of the auxiliary particles after $U$s are applied,
and the coefficients in the equation satisfy the normalization condition,
\begin{gather}
\label{conditions}
|\lambda_{00}|^2+|\lambda_{01}|^2 = 1,
\notag \\
|\lambda_{10}|^2+|\lambda_{11}|^2 = 1.
\end{gather}

When $U$ is performed on the decoy states $\left|+\right\rangle$ and $\left|-\right\rangle$ in $S^*$,
one can get
\begin{align}
\label{U-action-case-two}
U \left|+\right\rangle \left|\varepsilon\right\rangle 
& = \frac 1{\sqrt{2}}
( 
\lambda_{00} \left|0\right\rangle \left| \epsilon_{00} \right\rangle 
+ \lambda_{01} \left|1\right\rangle \left| \epsilon_{01} \right\rangle
+\lambda_{10} \left|0\right\rangle \left| \epsilon_{10} \right\rangle 
+ \lambda_{11} \left|1\right\rangle \left| \epsilon_{11} \right\rangle
)		
\notag \\
& 
=
\frac 12
\left|+\right\rangle	(
\lambda_{00} \left| \epsilon_{00} \right\rangle 
+ \lambda_{01} \left| \epsilon_{01} \right\rangle
+\lambda_{10} \left| \epsilon_{10} \right\rangle 
+ \lambda_{11} \left| \epsilon_{11} \right\rangle
)		
\notag \\
&
\quad+\frac 12
\left|-\right\rangle	(
\lambda_{00} \left| \epsilon_{00} \right\rangle 
- \lambda_{01} \left| \epsilon_{01} \right\rangle
+ \lambda_{10} \left| \epsilon_{10} \right\rangle 
- \lambda_{11} \left| \epsilon_{11} \right\rangle
),
\notag \\
U \left|-\right\rangle \left|\varepsilon\right\rangle & = \frac 1{\sqrt{2}}
( 
\lambda_{00} \left|0\right\rangle \left|\epsilon_{00}\right\rangle 
+ \lambda_{01} \left|1\right\rangle \left|\epsilon_{01}\right\rangle 
- \lambda_{10} \left|0\right\rangle \left| \epsilon_{10} \right\rangle 
- \lambda_{11} \left|1\right\rangle \left| \epsilon_{11} \right\rangle
)		\notag \\
&=\frac 12
\left|+\right\rangle	(
\lambda_{00} \left| \epsilon_{00} \right\rangle 
+ \lambda_{01} \left| \epsilon_{01} \right\rangle
-\lambda_{10} \left| \epsilon_{10} \right\rangle 
- \lambda_{11} \left| \epsilon_{11} \right\rangle
)		\notag \\
&\quad+\frac 12
\left|-\right\rangle	(
\lambda_{00} \left| \epsilon_{00} \right\rangle 
- \lambda_{01} \left| \epsilon_{01} \right\rangle
-\lambda_{10} \left| \epsilon_{10} \right\rangle 
+ \lambda_{11} \left| \epsilon_{11} \right\rangle
).
\end{align}
It can be seen from Eqs. \ref{U-action-case-one} and \ref{U-action-case-two} that if Eve does not want to be caught by 
the eavesdropping checking process in the protocol, $U$ must make the following equations hold:
\begin{gather}
\lambda_{01} = \lambda_{10} = 0,
\notag \\
\lambda_{00} \left| \epsilon_{00} \right\rangle = \lambda_{11} \left| \epsilon_{11} \right\rangle.
\label{conditions-for-two-level-protocol}
\end{gather}

Let us mark the Bell states used as information carriers in the protocol by
\begin{align}
\label{Bell}
\left| \psi^{\pm} \right\rangle = \frac 1{\sqrt{2}}
\left( \left|0b\right\rangle \pm \left|1\bar{b}\right\rangle \right),
\end{align}
where $b \in \{0,1\}$ and the bar over a bit value indicates its logical negation.
From Eqs. \ref{U-action-case-one}, \ref{U-action-case-two}, \ref{conditions-for-two-level-protocol}, 
when $U$ is performed on the single-particle state $\left\vert a \right\rangle$ where $a \in \{0,1\} $, 
we have
\begin{align}
\label{single}
U \left|a\right\rangle \left|\varepsilon\right\rangle = 
&\lambda_{a0} \left|0\right\rangle \left\vert \epsilon_{a0} \right\rangle 
+ \lambda_{a1} \left|1\right\rangle \left\vert \epsilon_{a1} \right\rangle		\notag \\
=&\begin{cases}
\lambda_{00} \left\vert 0 \right\rangle \left\vert \epsilon_{00} \right\rangle, & \text{if } a = 0;	\\
\lambda_{11} \left\vert 1 \right\rangle \left\vert \epsilon_{11} \right\rangle, & \text{if } a = 1,
\end{cases}	
\notag \\
= & \lambda_{aa} \left|a\right\rangle \left| \epsilon_{aa} \right\rangle,
\end{align}
then we can arrive at
\begin{align}
\label{fomerBell}
U_1 \otimes  U_2 \left|0b\right\rangle
\left|\varepsilon\right\rangle_1 \left|\varepsilon\right\rangle_2
= U_1 \left|0\right\rangle \left|\varepsilon\right\rangle_1
\otimes U_2 \left|b\right\rangle \left|\varepsilon\right\rangle_2
= \lambda_{00} \left\vert 0 \right\rangle \left\vert \epsilon_{00} \right\rangle
\otimes \lambda_{bb} \left|b\right\rangle \left\vert \epsilon_{bb} \right\rangle
= \lambda_{00}	\lambda_{bb}
\left\vert 0b \right\rangle
\left\vert \epsilon_{00} \right\rangle \left\vert \epsilon_{bb} \right\rangle.
\end{align}
Similarly, we can get
\begin{align}
\label{laterBell}
U_1 \otimes  U_2 \left|0\bar{b}\right\rangle
\left\vert \varepsilon \right\rangle_1 \left\vert \varepsilon \right\rangle_2 
= & U_1 \left|0\right\rangle \left\vert \varepsilon \right\rangle_1
\otimes U_2 \left|\bar{b}\right\rangle \left|\varepsilon\right\rangle_2 
\notag \\
= & \lambda_{00} \left\vert 0 \right\rangle \left\vert \epsilon_{00} \right\rangle
\otimes \lambda_{\bar{b}\bar{b}} \left\vert b \right\rangle \left\vert \epsilon_{\bar{b}\bar{b}} \right\rangle
\notag \\
= & \lambda_{00}	\lambda_{\bar{b}\bar{b}}
\left\vert 0 \bar{b}\right\rangle
\left\vert \epsilon_{00} \right\rangle \left\vert \epsilon_{\bar{b}\bar{b}} \right\rangle,
\end{align}
Combining Eqs. \ref{fomerBell} and \ref{laterBell}, we can get
\begin{align}
\label{UBell}
U_1 \otimes U_2
\left\vert \psi^{\pm} \right\rangle
\left\vert \varepsilon \right\rangle_1 \left\vert \varepsilon \right\rangle_2
= & U_1 \otimes U_2
\frac 1{\sqrt{2}}
\left( \left\vert 0b \right\rangle + \left\vert 1\bar{b} \right\rangle \right)
\left\vert \varepsilon \right\rangle_1 \left\vert \varepsilon \right\rangle_2   \notag \\
= & \frac 1{\sqrt{2}}
\left(
U_1 \otimes U_2 \left|0b\right\rangle
\left|\varepsilon\right\rangle_1 \left|\varepsilon\right\rangle_2 + 
U_1 \otimes U_2
\left\vert 1 \bar{b}\right\rangle
\left|\varepsilon\right\rangle_1 \left|\varepsilon\right\rangle_2
\right) \notag \\
= & \frac 1{\sqrt{2}}
\left(
\lambda_{00}	\lambda_{bb}
\left\vert 0b \right\rangle
\left\vert \epsilon_{00} \right\rangle \left\vert \epsilon_{bb} \right\rangle
+
\lambda_{00}	\lambda_{bb}
\left\vert 1 \bar{b}\right\rangle
\left\vert \epsilon_{00} \right\rangle \left| \epsilon_{bb} \right\rangle
\right)
\notag \\
= &
\lambda_{00}	\lambda_{bb}
\left\vert \psi^{\pm} \right\rangle
\left\vert \epsilon_{00} \right\rangle \left\vert \epsilon_{bb} \right\rangle.
\end{align}
It can be seen that no error will be introduced iff $\left\vert \psi^{\pm} \right\rangle$ and the ancillary particles are in a product state.
In this case, Eve cannot steal any useful information.
Therefore, entanglement-measurement attack can be defeated by eavesdropping checking using decoy photons.
Note here that we assume that Eve's entanglement operations act on all particles in the Bell state at the same time; 
When the entanglement operations only act on one particle of the Bell state,
it is easy to verify that $\left\vert \psi^{\pm} \right\rangle$ and the auxiliary particle are still in a product state, 
which will not be repeated here.

Next, let us consider the security proof for d-level-system-based quantum cryptography
against the entanglement-measurement attack. As mentioned earlier, such protocols mostly use
$\{ \left\vert k \right\rangle \}_{k=0}^{d-1}$ and $\{ F \left\vert k \right\rangle \}_{k=0}^{d-1}$ to perform eavesdropping checking.
We assume that the maximally entangled state used as the information carriers in the protocol is
d-level n-particle GHZ state,
\begin{align}
\label{GHZ-state}
	\left\vert \psi_{GHZ} \right\rangle =
	\frac{1}{\sqrt d} \sum_{j=0}^{d-1} \left\vert j,j,\dots,j \right\rangle_{1,2,\cdots,n}.
\end{align}
By the way, by setting $d=2$ and $n=3$, one can get the familiar three-particle GHZ state,
which is an entangled state frequently adopted in quantum cryptography protocols.

Suppose Alice sends some or all particles in each GHZ state $\left\vert \psi_{GHZ} \right\rangle$ 
with decoy particles to Bob in the protocol,
and Eve intercepts these particles. In this case, Eve's entanglement operation $U$ can be expressed as
\begin{align}
\label{U-definition-for-d-level-system}
U \left\vert l \right\rangle \left\vert \varepsilon \right\rangle 
=
\sum_{k=0}^{d-1} \lambda_{l,k} \left\vert k \right\rangle \left\vert \epsilon_{l,k} \right\rangle, 
\quad l = 0,1,2,\dots,d-1.
\end{align}
When $U$ is performed on the decoy states $\{ F \left| k \right\rangle \}_{k=0}^{d-1}\}$, we can arrive at
\begin{align}
\label{proof_Fd}
\boldsymbol{U} F \left\vert k \right\rangle \left|\varepsilon\right\rangle
= & \boldsymbol{U} \frac 1 {\sqrt{d}}
\sum_{l=0}^{d-1} \zeta^{lk} \left\vert l \right\rangle \left\vert \varepsilon \right\rangle
\notag \\
= & \frac 1{\sqrt{d}}
\sum_{l=0}^{d-1} \sum_{i=0}^{d-1} \sum_{j=0}^{d-1} \zeta^{lk}
\lambda_{i,j} \left\vert j \right\rangle \left\vert \epsilon_{i,j} \right\rangle.
\end{align}
Combining with the inverse Fourier transform formula
\begin{align}
\label{Fourier_inverse_transform}
\left\vert l \right\rangle = \frac {1}{\sqrt{d}} \sum_{k=0}^{d-1} \zeta^{-kl} F \left\vert k \right\rangle, 
\quad 
l \in \{ 0,1,2,\cdots,d-1 \},
\end{align}
we can get
\begin{align}
\label{transformation_d}
\boldsymbol{U} F \left\vert k \right\rangle \left|\varepsilon\right\rangle
= \frac 1d
\sum_{l=0}^{d-1} \sum_{i=0}^{d-1} \sum_{j=0}^{d-1} \zeta^{lk}
\lambda_{i,j} \left\vert j \right\rangle F \left\vert k \right\rangle \left\vert \epsilon_{i,j} \right\rangle.
\end{align}

As before, it can be seen from Eq. \ref{U-definition-for-d-level-system} that 
if Eve does not want her eavesdropping behavior to be discovered,
her attack cannot cause errors. In this case, her entanglement operation must satisfy
\begin{align}
\label{conclusion_1}
U \left\vert l \right\rangle \left|\varepsilon\right\rangle =
\lambda_{l,l} \left|l\right\rangle \left| \epsilon_{l,l} \right\rangle,
\end{align}
which means that
\begin{align}
\lambda_{l,l'} = \delta_{l,l'} \lambda_{l,l'}.
\end{align}
Similarly, from Eq. \ref{transformation_d}, Eve's entanglement operations also need to meet the following condition:
\begin{align}
\label{U-condition-2}
\sum_{l=0}^{d-1} \lambda_{l,l} \zeta^{(k-j)l} \left\vert \epsilon_{l,l} \right\rangle = 0, \quad k, l = 0,1,2,\dots,d-1.
\end{align}
When $k=0$, we can construct the matrix
\begin{align}
A =
\begin{bmatrix} 
1 & \zeta^{-1} & \zeta^{-2} & \zeta^{-3} & \cdots & \zeta^{-(d-1)}  \\ 
1 & \zeta^{-2} & \zeta^{-4} & \zeta^{-6} & \cdots & \zeta^{-2(d-1)}  \\
1 & \zeta^{-3} & \zeta^{-6} & \zeta^{-9} & \cdots & \zeta^{--3(d-1)} \\
\vdots & \vdots & \vdots & \vdots & \ddots  & \vdots \\
1 & \zeta^{-(d-1)} & \zeta^{-2(d-1)} & \zeta^{-3(d-1)} & \cdots & \zeta^{-(d-1)^2}
\end{bmatrix}_{(d-1) \times d},
\end{align}
One can get
\begin{align}
\label{rank-A}
rank (A) \le d-1.
\end{align}
Let us set
\begin{align}
B=
\begin{bmatrix} 
\zeta^{-1} & \zeta^{-2} & \zeta^{-3} & \cdots & \zeta^{-(d-1)}  \\ 
\zeta^{-2} & \zeta^{-4} & \zeta^{-6} & \cdots & \zeta^{-2(d-1)}  \\
\zeta^{-3} & \zeta^{-6} & \zeta^{-9} & \cdots & \zeta^{-3(d-1)} \\
\vdots & \vdots & \vdots & \ddots  & \vdots \\
\zeta^{-(d-1)} & \zeta^{-2(d-1)} & \zeta^{-3(d-1)} & \cdots & \zeta^{-(d-1)^2}
\end{bmatrix}_{(d-1) \times (d-1)},
\end{align}
then we can arrive at 
\begin{align}
\begin{vmatrix} 
B
\end{vmatrix} 
&=
\begin{vmatrix} 
1 & 1 & 1 & \cdots & 1  \\ 
\zeta^{-1} & \zeta^{-2} & \zeta^{-3} & \cdots & \zeta^{-(d-1)}  \\
\zeta^{-2} & \zeta^{-4} & \zeta^{-6} & \cdots & \zeta^{-2(d-1)} \\
\vdots & \vdots & \vdots & \ddots  & \vdots \\
\zeta^{-(d-1)+1} & \zeta^{-2(d-1)+2} & \zeta^{-3(d-1)+3} & \cdots & \zeta^{-(d-1)^2+(d-1)}
\end{vmatrix}
\times
\begin{vmatrix} 
\zeta^{-1} & 0 & 0 & \cdots & 0  \\ 
0 & \zeta^{-2} & 0 & \cdots & 0  \\
0 & 0 & \zeta^{-3} & \cdots & 0 \\
\vdots & \vdots & \vdots & \ddots  & \vdots \\
0 & 0 & 0 & \cdots & \zeta^{-(d-1)}
\end{vmatrix}
\notag \\
&= \zeta^{-\frac {d(d-1)}{2}} \prod_{1 \le j < i \le d-1} ( \zeta^i - \zeta^j ),
\end{align}
Obviously, 
$\begin{vmatrix} B \end{vmatrix} \ne 0$, 
thus $rank(B) = d-1$. Since $rank (A) \ge rank(B)$, $rank(A) \ge d-1$.
Combined Eq. \ref{rank-A}, easy to know $rank(A) = d-1$, 
which means that the homogeneous linear equations
\begin{align}
\label{U-condition-2}
\sum_{l=0}^{d-1} \lambda_{l,l} \zeta^{-jl} \left\vert \epsilon_{l,l} \right\rangle = 0, \quad l = 0,1,2,\dots,d-1,
\end{align}
only have a general solution. It is easy to see that the general solution is
\begin{equation}
\label{general-solution}
\lambda_{0,0} \left| \epsilon_{0,0} \right\rangle =
\lambda_{1,1} \left| \epsilon_{1,1} \right\rangle =
\lambda_{2,2} \left| \epsilon_{2,2} \right\rangle = \cdots =
\lambda_{d-1,d-1} \left| \epsilon_{d-1,d-1} \right\rangle.
\end{equation}

As before, we still assume that Eve's entanglement operations act on all particles in $\left\vert \psi_{GHZ} \right\rangle$.
According to Eqs. \ref{conclusion_1} and \ref{general-solution}, we can get
\begin{align}
\label{attack_GHZ}
& U_1 \otimes U_2 \otimes \cdots \otimes U_n \left| \psi_{GHZ} \right\rangle
\left\vert \varepsilon \right\rangle_0 \left|\varepsilon\right\rangle_1 \otimes \cdots \otimes
\left\vert \varepsilon \right\rangle_n
\notag \\
& =\frac{1}{\sqrt d} \sum_{j=0}^{d-1} U_1 \otimes U_2 \otimes \cdots \otimes U_n
\left\vert j,j,\dots,j \right\rangle_{1,2,\cdots,n}
\left\vert \varepsilon \right\rangle_0 \left|\varepsilon\right\rangle_1 \otimes \cdots \otimes
\left\vert \varepsilon \right\rangle_n
\notag \\
& =\frac{1}{\sqrt d} \sum_{j=0}^{d-1} \lambda_{jj} \lambda_{jj} \cdots \lambda_{jj}
\left\vert j,j,\dots,j \right\rangle_{1,2,\cdots,n}
\left\vert \epsilon_{jj} \right\rangle \left|\epsilon_{jj}\right\rangle \cdots
\left\vert \epsilon_{jj} \right\rangle  
\notag \\
& =\lambda_{00} \lambda_{00} \cdots \lambda_{00}
\frac{1}{\sqrt d} \sum_{j=0}^{d-1}\left| j,j,\dots,j \right\rangle_{1,2,\cdots,n}
\left\vert \epsilon_{00} \right\rangle \left\vert \epsilon_{00} \right\rangle \cdots
\left\vert \epsilon_{00} \right\rangle
\notag \\
& =\lambda_{00} \lambda_{00} \cdots \lambda_{00}
\left\vert \psi_{GHZ} \right\rangle
\left\vert \epsilon_{00} \right\rangle \left\vert \epsilon_{00} \right\rangle \cdots
\left\vert \epsilon_{00} \right\rangle,
\end{align}
It can be seen that $\left\vert \psi_{GHZ} \right\rangle$ and the ancillary particles are in a product state.
It is easy to verify that when the entanglement operation only acts on some particles in $\left\vert \psi_{GHZ} \right\rangle$, 
$\left\vert \psi_{GHZ} \right\rangle$ and auxiliary particles are still in a product state, which will not be repeated here.
Therefore, Eve cannot steal any useful information.
In conclusion, eavesdropping checking using decoy photons can effectively resist the entanglement-measurement attack.


\subsection{Security proof of using entangled states for eavesdropping checking}

As before, let us still assume that the quantum states used as information carriers 
are the d-level n-particle GHZ states shown in Eq. \ref{GHZ-state}.
Moreover, for simplicity, we directly use the entangled states for eavesdropping checking.

To achieve eavesdropping checking, Alice prepares some additional ones 
(assuming that the number is enough for eavesdropping checking)
when generating the GHZ state $\left\vert \psi_{GHZ} \right\rangle$.
Then she sends some or all of the particles in each state to Bob.
Afterwards, Alice and Bob negotiate (usually using classical channels),
and jointly randomly select some of the GHZ states and perform a quantum Fourier transform on 
each particle in each selected state. After the quantum Fourier transform, 
$\left\vert \psi_{GHZ} \right\rangle$ are transformed into
\begin{equation}
	\left\vert \psi_{GHZ}' \right\rangle =
	\frac{1}{d^{(d+1)/2}} \sum_{\sum_{i=1}^n j_i' \phantom{i} \textrm{mod} \phantom{i} d = 0}
	\left| j_1',j_2',\dots,j_n' \right\rangle_{1,2,\dots,n}.
\end{equation}
Finally, Alice and Bob perform single-particle measurements on each particle in $\left\vert \psi_{GHZ}' \right\rangle$.
At the same time, they randomly select some of the states $\left\vert \psi_{GHZ} \right\rangle$ 
together and perform single-particle measurements. For each quantum state, if the measurement results are denoted as
$\left\vert b_1 \right\rangle,\left\vert b_2 \right\rangle,\left\vert b_3 \right\rangle,\dots,\left\vert b_n \right\rangle$, then we have
\begin{align}
\label{GHZ-conditions}
\begin{cases}
b_1 = b_2 = \cdots = b_n, & \text{if the selected state is} \left|\psi_{GHZ}\right\rangle;	\\
\sum_{i=1}^n b_i \phantom{i} \textrm{mod} \phantom{i} d = 0, & \text{if the selected state is} \left|\psi_{GHZ}'\right\rangle.
\end{cases}
\end{align}

Alice and Bob can judge whether there are eavesdroppers in the quantum channel by checking 
whether the measurement results meet the conditions shown in Eq. \ref{GHZ-conditions}.
Let us now demonstrate how this eavesdropping checking method can resist entanglement-measurement attack.
Suppose Eve's entanglement operation is
\begin{align}
\label{GHZ-entanglement-operations}
\left\vert \Psi \right\rangle = U \left\vert \psi_{GHZ} \right\rangle \left\vert \varepsilon \right\rangle 
= \frac{1}{\sqrt d} \sum_{j=0}^{d-1} \left| j,j,\dots,j \right\rangle_{1,2,\cdots,n} \left| \epsilon_{j,j,\dots,j} \right\rangle,
\end{align}
where $\left\vert \varepsilon \right\rangle$ represents the auxiliary particle.
Performing quantum Fourier transforms on $\left\vert \psi_{GHZ} \right\rangle$, we can arrive at
\begin{align}
\label{QFT-GHZ}
\left|\varPsi_F\right\rangle 
& = F \otimes F \otimes \cdots \otimes F \left|\psi_{GHZ}\right\rangle \left| \epsilon_{j,j,\dots,j} \right\rangle \notag \\
& = \frac{1}{\sqrt d} \Big(
F \left|0\right\rangle F \left|0\right\rangle \cdots F \left|0\right\rangle \left| \epsilon_{0,0,\dots,0} \right\rangle 
+ F \left|1\right\rangle F \left|1\right\rangle \cdots F \left|1\right\rangle \left| \epsilon_{1,1,\dots,1} \right\rangle
+ F \left|2\right\rangle F \left|2\right\rangle \cdots F \left|2\right\rangle \left| \epsilon_{2,2,\dots,2} \right\rangle
+\cdots \notag \\
& \quad + F \left|d-1\right\rangle F \left|d-1\right\rangle \cdots F \left|d-1\right\rangle \left| \epsilon_{d-1,d-1,\dots,d-1} \right\rangle
\Big) \notag \\
& = \frac{1}{d \sqrt d} \Big(
\sum_{r_1^0, r_2^0, \dots, r_n^0 = 0}^{d-1} \left|r_1^0, r_2^0, \dots, r_n^0\right\rangle \left| \epsilon_{00\dots0} \right\rangle 
+
\sum_{r_1^1, r_2^1, \dots, r_n^1 = 0}^{d-1} \zeta^{\sum_{i=1}^n r_i^1}  
\left| r_1^1, r_2^1, \dots, r_n^1\right\rangle \left| \epsilon_{11\dots1} \right\rangle \notag \\
& \quad + 
\sum_{r_1^2, r_2^2, \dots, r_n^2 = 0}^{d-1} \zeta^{\sum_{i=1}^n r_i^2}  
\left| r_1^2, r_2^2, \dots, r_n^2\right\rangle \left| \epsilon_{11\dots1} \right\rangle
+\cdots + \sum_{r_1^{d-1}, r_2^{d-1}, \dots, r_n^{d-1} = 0}^{d-1} \zeta^{\sum_{i=1}^n r_i^{d-1}}  
\left| r_1^{d-1}, r_2^{d-1}, \dots, r_n^{d-1}\right\rangle \left| \epsilon_{11\dots1} \right\rangle
\Big) \notag \\
& = \frac{1}{d \sqrt d}
\left|00 \cdots 00\right\rangle \Big( 
\left| \epsilon_{00\dots0} \right\rangle + \left| \epsilon_{11\dots1} \right\rangle + 
\left| \epsilon_{22\dots2} \right\rangle + \cdots + \left| \epsilon_{d-1,d-1,\dots,d-1} \right\rangle
\Big) + \notag \\
& \quad \frac{1}{d \sqrt d} 
\left|00 \cdots 01\right\rangle 
\Big( 
\left| \epsilon_{00\dots0} \right\rangle + \zeta \left| \epsilon_{11\dots1} \right\rangle + 
\zeta^2 \left| \epsilon_{22\dots2} \right\rangle + \cdots + \zeta^{d-1} \left| \epsilon_{d-1,d-1,\dots,d-1} \right\rangle
\Big) + \notag \\
& \quad \frac{1}{d \sqrt d} \left|00 \cdots 02\right\rangle \Big( 
\left| \epsilon_{00\dots0} \right\rangle + \zeta^2 \left| \epsilon_{11\dots1} \right\rangle + 
\zeta^4 \left| \epsilon_{22\dots2} \right\rangle + \cdots + \zeta^{2(d-1)} \left| \epsilon_{d-1,d-1,\dots,d-1} \right\rangle
\Big) + \cdots + \notag \\
& \quad \frac{1}{d \sqrt d} 
\left|00 \cdots 0d-1\right\rangle 
\Big( 
\left| \epsilon_{00\dots0} \right\rangle + \zeta^{d-1} \left| \epsilon_{11\dots1} \right\rangle + 
\zeta^{2(d-1)} \left| \epsilon_{22\dots2} \right\rangle + \cdots + \zeta^{(d-1)^2} \left| \epsilon_{d-1,d-1,\dots,d-1} \right\rangle
\Big) + \notag \\
& \quad \frac{1}{d \sqrt d}
\left|00 \cdots 10\right\rangle \Big( 
\left| \epsilon_{00\dots0} \right\rangle + \left| \epsilon_{11\dots1} \right\rangle + 
\left| \epsilon_{22\dots2} \right\rangle + \cdots + \left| \epsilon_{d-1,d-1,\dots,d-1} \right\rangle
\Big) + \notag \\
& \quad \frac{1}{d \sqrt d} 
\left|00 \cdots 20\right\rangle 
\Big( 
\left| \epsilon_{00\dots0} \right\rangle + \zeta \left| \epsilon_{11\dots1} \right\rangle + 
\zeta^2 \left| \epsilon_{22\dots2} \right\rangle + \cdots + \zeta^{d-1} \left| \epsilon_{d-1,d-1,\dots,d-1} \right\rangle
\Big) + \notag \\
& \quad \frac{1}{d \sqrt d} 
\left|00 \cdots 30\right\rangle 
\Big( 
\left| \epsilon_{00\dots0} \right\rangle + \zeta^2 \left| \epsilon_{11\dots1} \right\rangle + 
\zeta^4 \left| \epsilon_{22\dots2} \right\rangle + \cdots + \zeta^{2(d-1)} \left| \epsilon_{d-1,d-1,\dots,d-1} \right\rangle
\Big) + \cdots + \notag \\
& \quad \frac{1}{d \sqrt d} 
\left|00 \cdots d-10\right\rangle 
\Big( 
\left| \epsilon_{00\dots0} \right\rangle + \zeta^{d-1} \left| \epsilon_{11\dots1} \right\rangle + 
\zeta^{2(d-1)} \left| \epsilon_{22\dots2} \right\rangle + \cdots + \zeta^{(d-1)^2} \left| \epsilon_{d-1,d-1,\dots,d-1} \right\rangle
\Big) + \cdots + \notag \\
& \quad \frac{1}{d \sqrt d} 
\left|d-1,d-1,\cdots,d-1\right\rangle 
\Big( 
\left| \epsilon_{00\dots0} \right\rangle + \zeta^{n(d-1)} \left| \epsilon_{11\dots1} \right\rangle + 
\zeta^{2n(d-1)} \left| \epsilon_{22\dots2} \right\rangle + \cdots + \zeta^{n(d-1)^2} \left| \epsilon_{d-1,d-1,\dots,d-1} \right\rangle
\Big).
\end{align}
From Eq. \ref{QFT-GHZ}, if Eve wants his attack not to be detected in eavesdropping checking, 
all items in the formula that do not meet the condition of $\sum_{i=1}^n b_i \phantom{i} \textrm{mod} \phantom{i} d = 0$ are equal to 0.
Therefore, it is easy to construct the matrix
\begin{align}
A =
\begin{bmatrix} 
1 & \zeta^{1} & \zeta^{2} & \zeta^{3} & \cdots & \zeta^{d-1}  \\ 
1 & \zeta^{2} & \zeta^{4} & \zeta^{6} & \cdots & \zeta^{2(d-1)}  \\
1 & \zeta^{3} & \zeta^{6} & \zeta^{9} & \cdots & \zeta^{3(d-1)} \\
\vdots & \vdots & \vdots & \vdots & \ddots  & \vdots \\
1 & \zeta^{d-1} & \zeta^{2(d-1)} & \zeta^{3(d-1)} & \cdots & \zeta^{(d-1)^2}
\end{bmatrix}_{(d-1) \times d},
\end{align}
As before, we have $rank(A) \le d-1$. Let us set
\begin{align}
B=
\begin{bmatrix} 
\zeta^{1} & \zeta^{2} & \zeta^{3} & \cdots & \zeta^{d-1}  \\ 
\zeta^{2} & \zeta^{4} & \zeta^{6} & \cdots & \zeta^{2(d-1)}  \\
\zeta^{3} & \zeta^{6} & \zeta^{9} & \cdots & \zeta^{3(d-1)} \\
\vdots & \vdots & \vdots & \ddots  & \vdots \\
\zeta^{d-1} & \zeta^{2(d-1)} & \zeta^{3(d-1)} & \cdots & \zeta^{(d-1)^2}
\end{bmatrix}_{(d-1) \times (d-1)},
\end{align}
Then we can get
\begin{align}
\begin{vmatrix} 
B
\end{vmatrix} 
&=
\begin{vmatrix} 
1 & 1 & 1 & \cdots & 1  \\ 
\zeta^{1} & \zeta^{2} & \zeta^{3} & \cdots & \zeta^{d-1}  \\
\zeta^{2} & \zeta^{4} & \zeta^{6} & \cdots & \zeta^{2(d-1)} \\
\vdots & \vdots & \vdots & \ddots  & \vdots \\
\zeta^{d-1+1} & \zeta^{2(d-1)+2} & \zeta^{3(d-1)+3} & \cdots & \zeta^{(d-1)^2+(d-1)}
\end{vmatrix}
\times
\begin{vmatrix} 
\zeta^{1} & 0 & 0 & \cdots & 0  \\ 
0 & \zeta^{2} & 0 & \cdots & 0  \\
0 & 0 & \zeta^{3} & \cdots & 0 \\
\vdots & \vdots & \vdots & \ddots  & \vdots \\
0 & 0 & 0 & \cdots & \zeta^{d-1}
\end{vmatrix}
\notag \\
&= \zeta^{\frac {d(d-1)}{2}} \prod_{1 \le j < i \le d-1} ( \zeta^i - \zeta^j ).
\end{align}
Therefore, 
$\begin{vmatrix} 
B
\end{vmatrix}
\ne 0
$ and $rank(A) = d-1$.
As before, we can get
\begin{align}
\left| \epsilon_{00\dots0} \right\rangle =
\left| \epsilon_{11\dots1} \right\rangle =
\left| \epsilon_{22\dots2} \right\rangle = \cdots =
\left| \epsilon_{d-1,d-1,\dots,d-1} \right\rangle.
\end{align}
Substituting this equation into \ref{GHZ-entanglement-operations}, we can get
\begin{align}
\left\vert \Psi \right\rangle
= \frac{1}{\sqrt d} \sum_{j=0}^{d-1} \left\vert j,j,\dots,j \right\rangle_{1,2,\cdots,n} \left\vert \epsilon_{00\dots0} \right\rangle
= \left\vert \psi_{GHZ} \right\rangle \left\vert \epsilon_{00\dots0} \right\rangle,
\end{align}
which indicates that $\left\vert \psi_{GHZ} \right\rangle$ and the auxiliary particles are in a product state.
Therefore, Eve cannot get any useful information by observing the auxiliary particles,
which means that her entanglement-measurement attack cannot succeed.


\section{Conclusion}

\noindent
As one of the main attacks against quantum cryptography, 
the entanglement-measurement attack has been playing an important role in the 
design and security analysis of quantum cryptography protocols.
We have provided the security proof for resisting this attack using two different eavesdropping checking methods.
The two eavesdropping checking methods are realized by single-particle states and entangled states respectively, 
and the protocol participants can choose according to the quantum devices they hold.
Our work has important guiding significance for the security analysis of quantum cryptography protocols, 
and will positively contribute to the development of quantum cryptography.


\end{document}